\documentstyle[aps,prb,epsfig,psfig,epsf]{revtex}
\begin{document}
\setlength{\parindent}{0em}
\draft
\title{Studies of the phase diagram of randomly interacting fermionic 
systems} 
\author{R. Oppermann$^{1,3}$ and B. Rosenow$^2$}
\address{$^1$Service de physique th\'eorique, CE Saclay, F--1191 
Gif--sur--Yvette, 
{\it France}\\
$^2$ MPI f\"ur Kernphysik, D--69029 Heidelberg, {\it Federal Republic of
Germany}\\
$^3$ Institut f. Theoret. Physik, Univ. W\"urzburg D--97074
W\"urzburg, {\it Federal Republic of Germany}}
\date{\today}
\maketitle
We present details of the phase diagrams of fermionic systems with random
and frustrated interactions emphasizing the important role of the
chemical potential $\mu$. 
The insulating fermionic Ising spin glass model is shown to reveal 
different entangled magnetic instabilities and phase transitions.
We review tricritical phenomena related to the strong correspondence 
between charge and spin fluctuations and controlled by quantum statistics.
A comparison with the diluted Sherrington--Kirkpatrick spin glass and with 
classical spin 1 models such as the Blume--Emery--Griffiths model is presented. 
We present a detailed analysis for the infinite range model showing that
spin glass order must  decay discontinuously as $\mu$ exceeds a 
critical value, provided the temperature is below the tricritical one, and 
that the zero temperature transition is of classical type. 
Replica permutation symmetry breaking (RPSB) of the Parisi type 
describes the thermal spin glass transitions, along with 
modifications of the SK--models Almeida--Thouless--line.
RPSB occurs in any case on the irreversible side of the (fermionic) 
Almeida Thouless
lines and hence at least everywhere within a fermionic spin glass phase.
Although the critical field theory of the quantum paramagnet to spin
glass
transition in metallic systems
remains replica--symmetric at $T=0$, with only small corrections at low T
from RPSB, the phase diagram is
affected at $O(T^0)$ by RPSB. 
Generalizing our results for the fermionic Ising spin glass we consider
aspects of models with additional spin and charge quantum--dynamics
such as metallic spin glasses.


\pagebreak
\section{Introduction and summary}
In this paper we concentrate on the magnetic phase diagram of the fermionic
Ising spin glass and some of its related models, including a metallic 
extension. While in the preceding paper (denoted by I and the 
present 
one by II in the following) we showed the particular importance of Parisi 
replica symmetry breaking (RPSB) \cite{parisi1,parisi2} for the $T=0$-- and 
low--T properties of 
fermionic correlations we now consider mainly thermodynamic aspects. 
A fascinating and challenging selfconsistency problem shows up: knowledge 
of the phase diagrams is necessary to see where RPSB must be taken into 
account (no matter how complicated replica--symmetric calculations have 
already been) and finally the phase diagram, as paper I indicates, 
depends partially on RPSB. 
With respect to the absence of RPSB outside spin glass phases and in zero 
magnetic field, the quantum spin glass models of 
all colors are fortunately quite in the tradition of classical spin glasses 
\cite{binyg,FiHe,rammal}. This helps to gain information for and
from those outside areas without the complications of RPSB.\\
In II, we present details of the tricritical behavior observed recently in 
thermal 
and in quantum--critical transitions of both insulating and metallic 
model versions. A review of previous results for tricritical phenomena 
\cite{BRRO} is 
joined with new details obtained for the discontinuous regime at higher 
chemical potentials, for low temperature solutions in general, and for 
instabilities inside the spin glass phase which are not primarily linked to 
replica permutation symmetry breaking.\\            
Considering fully frustrated magnetic interactions it is natural to expect
spin glass order to play a major role in the phase diagram, and it will 
thus be surprising to find other types of transitions. Magnetic phenomena of 
these fermionic models, which are naturally described in the grand canonical 
ensemble, react strongly to the variation of the particle pressure 
controlled by the chemical potential. 
Charge and spin fluctuations are forced by quantum statistics to cooperate, 
which leads to the tricritical phenomena and to further instabilities 
in the paramagnetic regime as well as inside the spin glass phase. 
In addition to the conventional scenario of spin glass transitions (left 
aside the possibility of reentrant behavior)
an unconventional critical line, linked to the creation of 
additional metastable solutions, passes through the tricritical point and 
extends into the entire spin glass and paramagnetic phase. \\
While the analysis of fermionic spin glasses deals only with real chemical
potentials in first place, it is also clear that all so far considered 
quantum spin glasses (a prominent example being the transverse field Ising
spin glass) can be viewed as fermionic spin glasses with 
properly chosen imaginary chemical potentials. This choice 
and the phase diagram of the model, comparing for example spin 
$\frac{1}{2}$-- and spin 1--models, depends on the spin quantum number.
Thus it becomes evident that understanding the phase diagram of fermionic
spin glass models within the whole plane of complex chemical potentials
will be important. We note that this point of view also applies to the 
twodimensional Ising model, recalling the success of other representation 
in terms of Majorana fermions or what has been called Ising fermions  
\cite{Onsager,Itz1,Zuber}. 
The application to experiments which are described by fermionic
systems with random interaction is very likely a wide open field 
\cite{Mydosh}.
Yet a large number of experimental results already exists and we cite 
here a few \cite{mydosh,steglich,dai}. The natural field of 
application
includes High$T_c$superconductors, heavy fermion systems, semimagnetic 
semiconductors etcetera.\\ 
One must be concerned in general with low temperature behavior near ($T=0$) 
quantum phase transitions, which is very susceptible to quantum dynamical 
effects, but also interesting types of (asymptotically classical) thermal 
tricritical transitions which mix spin-- and charge--fluctuations occur.
Interference of spin glass features in charge response and transport 
properties \cite{Georges} is of central interest. \\
The symmetry classification of the concerned QPT's turned out to be very 
different from that of thermal phase transitions and also does not appear
to resemble, for example, the $T=0$ QPT universality classifications known 
for Anderson localization.\\
In early papers, using the spin--static approximation, the fundamental 
question had been raised \cite{goldschmidt,usabuett,thirumalai}
whether tunneling through energy barriers in contrast to thermal hopping over 
them might distinguish quantum spin glasses from classical ones.
The quantum--mechanical image of this type of built--up of zero 
temperature spin glass order could have been given by a Parisi order parameter 
function $q({\cal{E}},x)$ in place of the known classical one $q(T,x)$ with 
${\cal{E}}$ being a parameter driving the system at $T=0$ through the 
transition.
In the preceding paper, cited henceforth as paper I, we have shown that 
this is not the case. Instead the quantum dynamical image of Parisi 
replica symmetry breaking is extremely important even at $T=0$ but in a 
totally different way. It determines elementary features of the fermionic 
correlations. 
The strength of these effects as observed at half--filling and the 
qualitative changes entrained by them leave open the question in which
way the magnetic phase diagram is affected by Parisi RPSB. This is very 
hard to answer for the moment, since the analysis may require a 
sufficiently good approximation of the Parisi function at $T=0$ (a fact
that has not been necessary at half--filling). On the other hand our 
analysis of paper I did not show any sign of a limitation of RPSB-- 
effects to half--filling. On the contrary unsurmountable difficulties at 
finite order of RPSB in deriving stable homogeneous saddle--point 
solutions away from half--filling could possibly be cured by use of the full 
fermionic Parisi solution for low temperatures. This requires however the 
knowledge not only of the Parisi function $q(x)$ at low T 
\cite{parisi1,parisi2} but also of quantum--dynamic fermionic Green's 
functions as described in the preceding paper.\\
Quantum spin glasses form
indeed a link between the general statistical theory of spin glasses and
randomly interacting systems on one hand and of interacting many fermion 
systems on the other. Beyond this secondary role they also assume 
their unique place. There are clear--cut differences between for example
$T=0$ spin glass transitions and other quantum phase transitions, be they
magnetic or electronic like the Anderson localization. As for their thermal
transitions they can, despite some relationships, as well be clearly 
distinguished from random field systems for example. \\
Charge correlations and fluctuations in metallic spin glasses are of 
particular interest, since electronic transport uses the part of the 
Hilbert space which is spanned by nonmagnetic states. 
Magnetic transitions alter nonanalytically the occupation of these 
states which in
turn leads to nonanalytic charge fluctuations \cite{BRRO}. These secondary
critical phenomena comprise for example the effect of non--Fermi liquid
behavior in the vicinity of a quantum spin glass transition \cite{Georges}.
for metallic
quantum spin glass transitions had been given before \cite{SRO}, emphasizing 
similarities and differences with respect to 
the one for transverse field Ising spin glasses \cite{RSY}.  
Charge variables are now included to describe the corresponding fluctuations.


\pagebreak
\section{Tricritical phase diagram of an Ising spin glass with charge 
fluctuations}
One crucial feature of fermionic spin glasses is the intimate connection
of spin and charge degrees of freedom. Quantum statistics with the presence
of nonmagnetic states (single and double occupancy) opens the possibility
of  charge fluctuations which thermally dilute the spin system. 
The simplest model displaying this type of behavior is the  SK--model on a 
fermionic space with four states per site, denoted by $ISG_f$ and defined 
by the Hamiltonian
\begin{eqnarray}
H=-\frac{1}{2} \sum_{i\neq j} J_{ij} \sigma_i \sigma_j - \mu \sum_i n_i
\end{eqnarray} 
with
spins $\sigma_i =  \Psi^{\dagger}_{i, \alpha} \sigma^z_{\alpha \beta}
\Psi_{i, \beta}$ , particle number operator $n_i= \Psi^{\dagger}_{i, \alpha}
\Psi_{i, \alpha}$  and Gaussian distributed exchange integrals
$J_{i j}$ with variance $J^2$ .\cite{opgro}. The fermionic field operators 
obey the usual commutator
relations $ \{ \Psi_{i \alpha}, \Psi_{j \beta} \} =0$ and 
$ \{ \Psi^{\dagger}_{i \alpha}, \Psi_{j \beta} \} = \delta_{i j} \delta_
{\alpha \beta} $.
As a guide to the global phase diagram we study an exactly solvable
infinite range version of the model but the formulae obtained in this 
subsection may equally well be considered as the saddle point 
approximation for an interaction with finite range.
We recall the free energy (thermodynamic potential) for arbitrary order K
of Parisi replica symmetry breaking \cite{opgro}
\begin{eqnarray}
\beta f& =& \frac{1}{4} \beta^2 J^2 \left[(1-\tilde{q})^2 -(1-q_1)^2 +
q_1^2-\int_0^1  dx\, q^2(x)\right]-\ln 2 - \beta \mu
\label{m2one}\\
&-& \lim_{K \to \infty} \int_{z_{K+1}}^G \ln \left[ \int_{z_K}^G
\left[...\left[ \int_{z_1}^G \left( \cosh(\beta \tilde{H}) 
+\cosh(\beta \mu)\exp(-\frac{1}{2}\beta^2 J^2(\tilde{q}-q_1)) \right)^{
m_1} \right]^{\frac{m_2}{m_1}}\right. \right.
\left. \left.  ... \right]^{\frac{m_K}{m_{K-1}}}\right]^{\frac{1}{ 
m_K}}\nonumber
\end{eqnarray}
The abbreviating notation $\int_z^G$ with upper index G denote normalized 
Gaussian 
integrals over $z$, ie $\int_{-\infty}^{\infty} 
dz\frac{exp(-\frac{z^2}{2})}{\sqrt{2\pi}}$, and $q(x)$ coincides
with the Parisi order parameter function (the replica diagonal part
$\tilde{q}$ is not contained in it). The fields $z_{\nu} $ are 
needed
for decoupling the  $m_{\nu} \times m_{\nu}$ blocks of the Parisi
matrix Q and the effective field $\tilde{H}$ is the sum of the external field
$h$ and the decoupling fields $ J \sum_{\nu =1}^{K+1} \sqrt{q_{\nu} - 
q_{\nu +1}} z_\nu$.
Notice the appearance of the Parisi variables $q_\nu$ and the additional
$\tilde{q}$ which lies at the heart of the following discussion. The 
Edwards - Anderson order parameter $q_{EA} = \lim_{t \to \infty} <S_i(t)
S_i(0)>$ describes the freezing of spins in the spin glass phase and is 
given by the Plateau height
$q(1)$ of the order parameter function, 
whereas the replica diagonal $\tilde{q} = [< \sigma^a 
\sigma^a>]_{av}$ 
is related to the average filling factor $[\nu]_{av} = [<\Psi^{\dagger}_
{\alpha} \Psi_{\alpha}>]_{av }$ by 
$[\nu]_{av} =1+ \tanh(\beta \mu)(1 - \tilde{q})$. The last relation is exact 
even in the case of replica symmetry breaking.
\subsection{Spin glass transition and unusual critical line}
To understand the effects of charge fluctuations and to gain insight into 
the global phase diagram it is sufficient to 
consider a replica symmetric approximation of eq.(\ref{m2one}) with $q(x)=q$ 
constant
\begin{eqnarray}
f &=& 1/4 \beta [(1-\tilde{q})^2-(1-q)^2]- T\ln 2-\mu\label{two}
-T \int_z^G\ln\{{\cal{C}}_{\mu}(z)\},\\
{\cal C}_{\mu} (z) & =&    \cosh[\beta\tilde{H}(z)] +
\cosh(\beta\mu)\exp[-1/2\beta^2(\tilde{q}- q)]
\end{eqnarray}
where $\tilde{H}(z) $ stands for $ J \sqrt{q} z + h$. Differentiation with 
respect to the saddle point variables q and $\tilde{q}$ yields 
the corresponding selfconsistency equations
\begin{eqnarray}
q & = & \int_z^G \frac{\sinh^2[\beta \tilde{H}(z)]}{{\cal C}^2_{\mu}(z)}
\label{three}\\
\tilde{q}&=&\int_z^G\frac{cosh[\beta\tilde{H}(z)] }{{\cal C}_{\mu}(z)}
\label{four}
\end{eqnarray} 
whereas differentiation with respect to h and $\mu $ yields the equations for
the magnetization and the average filling $[\nu]_{av}$
\begin{eqnarray}
m & = & \int_z^G \frac{sinh(\beta \tilde{H}(z))}{{\cal C}_{\mu} (z)}
\label{six}\\[0cm]                 
 [ \nu ]_{av}& = & [ < \Psi^{\dagger}_{\sigma} \Psi_{\sigma}>]_{av}\\
& = & 1+\int_z^G \frac{\sinh(\beta \mu) e^{-\frac{1}{2} (\beta J)^2
(\tilde{q}- q)}}{{\cal C}_{\mu}(z)}
\label{seven}
\end{eqnarray}
Phase transitions are signalized by vanishing masses of the order parameter
propagators which in the saddle point formalism are given by second 
derivatives of the free energy. 
On the other hand a positive mass for $\tilde{q}$ and a 
negative one for $q$ guarantees stability. In the paramagnetic regime
the relevant expressions are
\begin{eqnarray}
\frac{\partial^2 f}{\partial \tilde{q}^2}=\frac{1}{2}\beta 
[1-1/2\beta^2\tilde{q}(1-\tilde{q})]&>&0 
\label{eight}\\
\frac{\partial^2 f}{\partial q^2}=\frac{1}{2}\beta[-1+\beta^2\tilde{q}^2]&<&0
\label{nine}
\end{eqnarray} 
This approach correctly tests critical fluctuations in the paramagnetic 
regime, in the following sections a more complete
analysis including replica symmetry breaking fluctuations
will be carried out using the approach of Almeida and Thouless \cite{AT}. 
A similar system of coupled stability conditions was found for the 
BEG - model \cite{BEG} and for a SK - model with crystal field
\cite{MS}, the case of half filling was considered in \cite{Theu}.
Analyzing (\ref{nine}) as an equation we get the condition $T_c=\tilde{q}_c$ 
for the transition to spin glass order and solve (\ref{four}) for
\begin{eqnarray}
\mu_{c1}(T)=T \cosh^{-1}[(1/T-1)\exp[1/(2T)]]
\label{ten}
\end{eqnarray}
In the same way we calculate
the curve corresponding to critical charge density fluctuations
\begin{equation}
\mu_{c2}(T)=T \cosh^{-1}[\frac{(1\mp\sqrt{1-8T^2})^2}{8T^2}
\exp[\frac{2}{1\mp\sqrt{1-8T^2}}]]
\label{eleven}
\end{equation}
The graphs of these two functions are displayed in figure 1; they cross
each other at $\mu=.958,T=.35$ and have a common tangent point at
 $\mu=1/3 \cosh^{-1}[2\exp(3/2)]=0.961056, T=1/3$.
There seems to be a region where the diagonal elements of the Q - matrix 
(e.g. charge fluctuations) become critical before the off - diagonal elements 
which may indicate some type of phase transition. 
To shed some light upon this problem it is instructive
to solve the selfconsistency equation (\ref{four}) for $q=0$. 
For $T\rightarrow\infty$
there clearly exists only the physically meaningful solution $\tilde{q}=1/2$.
The solution stays unique until one crosses (coming from large temperatures)
the $\mu_{c2}$ line the first time. At this temperature two new solutions 
show up
- one maximum and one minimum of the free energy. However, the physical solution
still has a lower free energy than the new solutions and in addition to 
that the new solutions violate the stability condition (\ref{nine}).
For $.95<\mu<.962$ one crosses the $\mu_{c2}$ line a second time when lowering
the temperature. 
Now something more severe happens: the maximum which at the first crossing
coincided with the second minimum moves to the physical minimum, merges
with it and then disappears. This doesn't have any consequences as the system
is in the ordered phase at this temperature anyway and playing with paramagnetic
solutions seems to be of academic interest only but there is one important 
exception:
the point $(\mu=.962, T=1/3)$. At this point the physical (paramagnetic) 
solution is unstable to both diagonal and offdiagonal fluctuations, a 
situation that deserves further investigation.
The second surprise is that the $\mu_{c1}(T)$ curve which at least for small
$\mu$ corresponds to a second order phase transition to a spin glass phase 
turns around at $\mu=.961056$ and thus leaves us with the question about the 
nature of the low temperature state of the system. To gain some insight into
this question we look at exact low temperature solutions to the selfconsistency
equations (\ref{three}) and (\ref{four}).

\begin{figure}
\epsfig{file=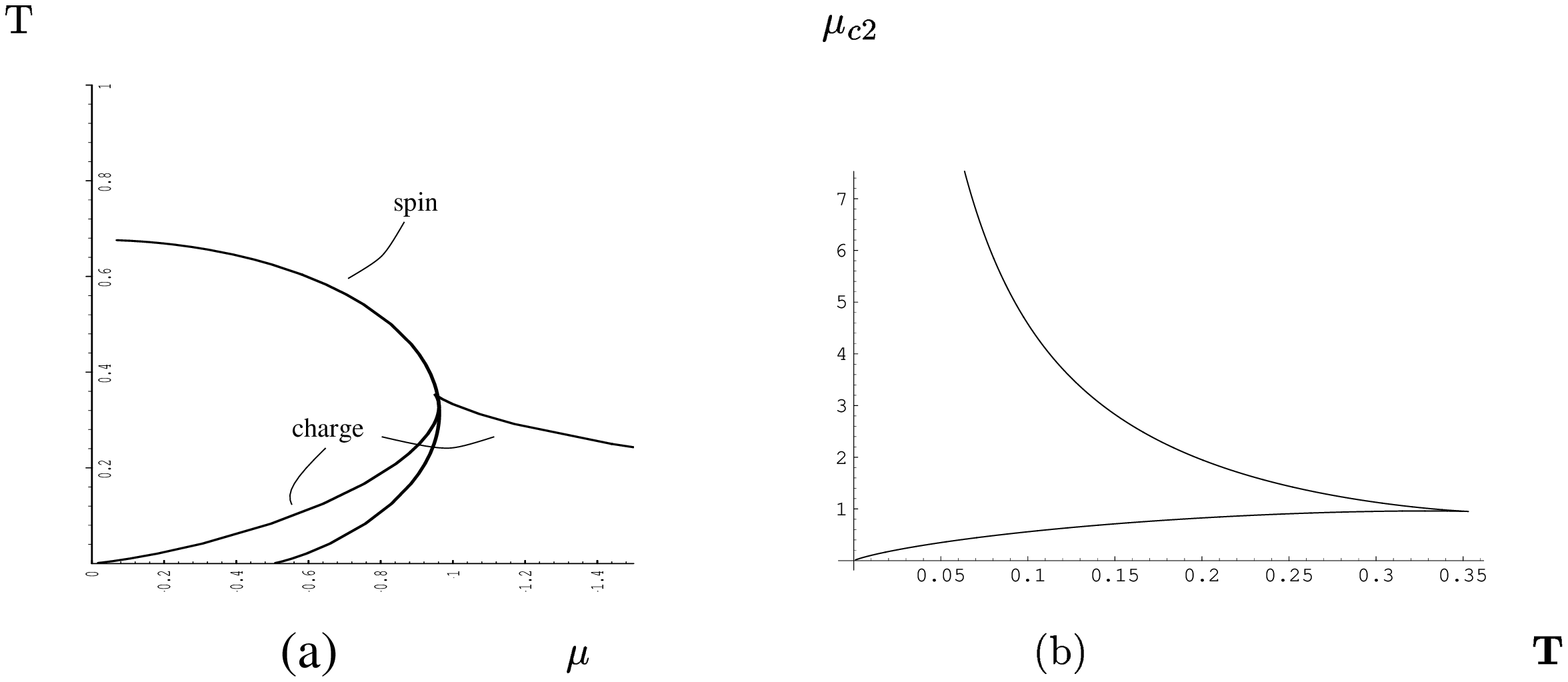,width=7.8cm,angle=270}
\caption{(a): Lines of diverging spin glass susceptibility, bending 
smoothly 
around from $(T=.6767,\mu=0)$ to $(T=0,\mu=.5)$, and unconventional line
$T_{c2}(\mu)$
of divergent replica diagonal nonlinear susceptibility $\chi^{aaaa}$.
(b) only the unconventional line is displayed 
here and given as $\mu_{c2}(T)$.}
\label{muc2}
\end{figure}

\pagebreak
\subsection{Exact low temperature solutions}
To analyze  solutions of the saddle point equations at low temperatures it
is useful to introduce the abbreviation $x = 
\cosh(\beta \mu)\exp[-1/2\beta^2(\tilde{q}-q)]$ and to  
integrate the difference of eqs. (\ref{three}) and (\ref{four})

\begin{eqnarray}
\tilde{q} - q&=& \int_y^G\frac{1+\cosh(\beta\sqrt{q}z) x}
{(\cosh(\beta\mu)+x)^2}
\label{twelve}\\
&=& \frac{T}{\sqrt{2\pi q}}\int_{-\infty}^{\infty}dy\frac{1+x \cosh(y)}
{(x+\cosh(y))^2} = O(T^3)\nonumber\\
& =& \frac{2 T}{\sqrt{2 \pi q}} + O(T^3)\ \ ,
\label{thirteen}
\end{eqnarray}

surprisingly enough the result is independent of x.\\
The leading low temperature contributions to the $\tilde{q}$-
equation can be integrated exactly. Its relevant asymptotic behavior, 
needed for the following discussion, is extracted by

\begin{eqnarray}
\tilde{q}&=&1-\frac{x T}{\sqrt{2 \pi q}}\int_{-\infty}^{\infty}
dy \frac{1}{\cosh(y) + x}
\label{fourteen}\\
&=&1-\frac{2 x T}{\sqrt{2 \pi q}}       
\left\{\begin{array}{c} \frac{\pi}{2}\; \; for\; x \to 0 \\
\frac{1}{x} \ln(2 x)\;\; for\; x \to \infty \end{array} \right.\nonumber
\end{eqnarray}

 The behavior of x in the low T limit is determined by the 
 sign of $ |\mu| - \beta/2 (\tilde{q}-q) = |\mu|-\frac{1}{\sqrt{2 \pi q}}$.
 For $|\mu| < \frac{1}{ \sqrt{2 \pi q}}$ x vanishes and we obtain 
 $\tilde{q}=1 -\sqrt{\frac{\pi}{8}} T
 \exp[\beta(|\mu|-\frac{1}{\sqrt{2 \pi q}})]$, $q=1-\sqrt{\frac{2}{\pi}} T$.
 For $\mu=\frac{1}{\sqrt{2 \pi}}$ x stays finite when  T goes to zero and 
 one still has $q(T=0)=\tilde{q}(T=0) =1$. For still larger values of the
 chemical potential this value decreases, since now x grows exponentially 
 and eq(\ref{fourteen}) turns into
 
\begin{eqnarray}
 q=\tilde{q}=1-\frac{2 x T}{2 \pi q} \frac{1}{x}\ln(x)
 = 1-\frac{2}{\sqrt{2 \pi q}}(|\mu|-\frac{1}{\sqrt{2 \pi q}}).
\label{fifteen}
 \end{eqnarray}
 
 The last equation is of fourth order in $\sqrt{q}$ and can have
 several
solutions. In the limit $\mu \to\infty$ however the solution becomes
unique
and an expansion in inverse powers of the 
 chemical potential seems useful. One finds 
\begin{eqnarray}
q=\frac{1}{2 \pi}(1/\mu^2+1/\mu^4 + O(1/\mu^6))\ .
\end{eqnarray}
 We now compare the free energy of the 
 paramagnetic solution with that of the ordered one.\\
Below the $\mu_{c2}$--line there exist three paramagnetic solutions,
two
minima and one maximum of the free energy. The maximum with the lower 
free energy ($\tilde{q}=1$) is unstable with respect to spin
glass
order, hence the physical solution is  
\begin{eqnarray}
\tilde{q}_{para} = 2 e^{-\beta \mu} + O(\beta^2 e^{-2 \beta \mu}).
\end{eqnarray}

The corresponding free energy at zero temperature is 
$f_{dis}=-2 \mu$, whereas  the free energy of the spin glass solution is

 \begin{eqnarray}
 f_{ordered}&=&\frac{1}{4}\beta[-2(\tilde{q}-q)+\tilde{q}^2-q^2]-\mu
\label{sixteen}\\
 & &-T\int_z^G[\ln(1+\frac{x}{\cosh(\beta \sqrt{q} z)})
 +\ln(\cosh(\beta \sqrt{q} z)]\nonumber\\
 &=&\frac{-1+q}{\sqrt{2 \pi q}} -\mu-\frac{2 \sqrt{q}}{2 \pi}
 \int_0^{\infty}dz\, z \exp(-\frac{z^2}{2})\nonumber\\
& &-\frac{2 T^2}{\sqrt{2 \pi q}}
 \int^x_0 d\tilde{x}\int_0^{\infty}dy\frac{1}{\tilde{x}+\cosh(y)}\nonumber\\
 &=&-2-\mu+\frac{1}{\mu}(1/2+\sqrt{\frac{q}{2 \pi}}-2 \sqrt{\frac{q}{2 \pi}}
 -1/4)\nonumber\\
&=&-2 \mu+\frac{1}{\mu}(\frac{1}{4}-\frac{1}{2 \pi})\nonumber
 \end{eqnarray}
Hence at least for large values of the chemical potential this replica
symmetric analysis indicates a paramagnetic  ground state.
On the other hand we know \cite{opgro} that for half filling the 
effect of the fermionic space is to lower the temperature of the continuous
spin glass transition by a factor of $0.6767$. 
Hence a transition from spin glass to paramagnet must take place on the 
zero temperature axis. In order to locate it
we solve Eq.(\ref{fifteen}) for all values of the chemical potential and 
compare the free energy of the different solutions. 
Spin glass order is realized up to a chemical potential
$\mu =.900$, at this value we find a first order transition with a jump of
the order parameter from $q=.603$ to zero. 
By means of the relation $[\nu]_{av} = 2 -\tilde{q}$ (valid in this form 
only at $T=0$) one finds phase separation for fillings $1.34 < \nu < 2$. \\
However, a stability analysis of this replica symmetric solution 
according to the 
scheme of de Almeida and Thouless \cite{AT,Brasil} reveals two negative 
eigenvalues of the Hessian matrix instead of the expected one negative value 
indicating just the instability towards replica symmetry breaking.
As described in \cite{binyg} for spin glass problems, one has to choose
the {\it stable} solution with the lowest free energy, hence the system 
undergoes a first order transition from half filling to 
the completely filled paramagnetic state at $\mu =
\frac{1}{\sqrt{2\pi}}J$.
The analysis of RPSB in I indicates that this region of
incompressibility
becomes smaller with increasing order of symmetry breaking. Therefor
one may expect that for infinite order RPSB  the fermion filling
increases
continuously from one with increasing chemical potential until it makes
a finite jump at a special value of $\mu$.\\
\begin{figure}[h!]
\epsfig{file=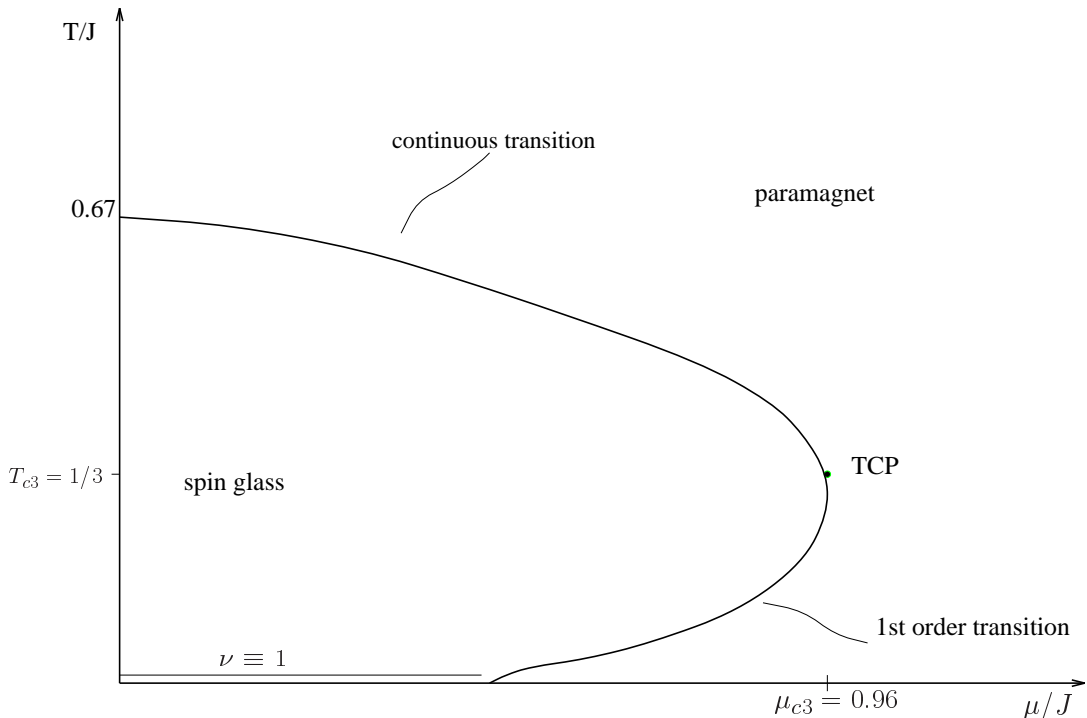,width=10 cm,angle=270}
\label{sepn}
\caption{Qualitative phase diagram of the $ISG_f$ in ($\mu$,T)--space 
calculated in the replica symmetric approximation}
\end{figure}

Now we have gathered important information on the global phase diagram
of the infinite--range fermionic spin glass model. 
At $\mu =0$ (half filling) a continuous transition takes place, for 
increasing $\mu$ the transition temperature is lowered.  
At zero temperature the transition is first order in both the spin and charge 
density, therefore it is natural to look for a tricritical point at which 
the transition changes its character from 2nd order to 1st order. 
The point $T=J/3, \mu =0.961056$
is a good candidate for tricritical behavior since charge-- and spin--
fluctuations become simultaneously critical there. 
\subsection{Tricritical Behavior}
Resorting to different techniques we performed a detailed analysis of 
tricritical behavior in the $ISG_f$. Mean field type quantities are most
easily calculated from an expansion of the saddle point free energy around 
the 
tricritical point given by
\begin{equation}
T_{c3}=J/3\quad {\rm and}\quad \mu_{c3} = \frac{J}{3} 
arcosh(2 \exp(\frac{3}{2})) \simeq .9611J
\end{equation}
This expansion reads

\begin{eqnarray}
f&-&f_{TCP}=\mu -\mu_{c3}-\frac{3 h^2}{2 J} (\delta\tilde{q}-\int_0^1 q(x)
dx)\nonumber\\
&+&J\{(\frac{3}{2}r_g 
g-r_{T}\delta T^2)\delta\tilde{q}
+\frac{3}{2}\delta T [(\delta\tilde{q})^2\nonumber\\ 
&-&\int^{1}_{0}dxq^2(x)]
-\frac{3}{2}
[\int^{1}_{0}dx[xq^3(x)+3q(x)\int^{x}_{0}dyq^2(y)]\nonumber\\
&-&3\delta\tilde{q}\int^{1}_{0}dxq^2(x)
+\frac{1}{4}(\delta\tilde{q})^3]-\frac{y_4}{4}\int^{1}_{0}dxq^4(x)\} ,
\label{seventeen} 
\end{eqnarray}

where $\delta\tilde{q}\equiv\tilde{q}-\tilde{q}_{TCP}, 
gJ=\mu-\mu_{c3 }+ 3 (\zeta^{-1}J-\mu_{c3})\delta T$ as nonordering
field \cite{LawSa}, and 
$\delta T \equiv T-T_{c3}$. The constants are given by 
$r_g=\frac{2\zeta}{3}, 
r_{T}=12(1-\frac{3}{4}\zeta^{-2})$ 
with $\zeta \equiv tanh(\mu_{c3}/T_{c3})\simeq 0.9938$, and 
$\mu_{c3}$
as the 
characteristic chemical potential locating the {\it TCP}.
The average filling factor corresponding to $\mu_{c3}$ is evaluated
as $[\nu_{c3}]_{av}\simeq 1.6625$. 
Moreover, we derived a fluctuation Lagrangian for the tricritical and finite 
range
$ISG_f$; a Lagrangian of the same structure is obtained for generalized
models (eg with a transport mechanism) at finite temperature by integrating
out dynamical degrees of freedom

\begin{eqnarray}
L&=& \frac{1}{t}\int d^dx{\large[}
-\frac{3}{2} h^2 \sum Q^{ab}\nonumber\\
& +&\frac{r\kappa_1}{(\kappa_2)^2}\sum Q^{aa} 
+ \frac{1}{2}\sum Q^{aa}(-\nabla^2+u)Q^{aa}
\nonumber\\
&+&\frac{1}{2}Tr^{\prime}(\nabla Q^{ab})^2 
- \frac{1}{t}\sum^{\hspace{.6cm}\prime} Q^{aa}Q^{bb}  
-\frac{\kappa_1}{3}\sum (Q^{aa})^3\nonumber\\
&-&\frac{\kappa_3}{3}
Tr^{\prime}Q^3
-\kappa_2\sum^{\hspace{.5cm}\prime} 
Q^{aa}Q^{ab}Q^{ba}+\frac{y_4}{4}\sum^{\hspace{.5cm}\prime} (Q^{ab})^4 
{\large],} 
\label{eighteen}
\end{eqnarray} 

Here 
$4(\frac{\kappa_1}{t})^{(0)}=(\frac{\kappa_2}{t})^{(0)}=
(\frac{\kappa_3}{t})^{(0)}
=\frac{3^3}{2}$ 
and $u^{(0)}=0$ denote the bare coefficients at tricriticality.  
One fourth order term relevant for replica symmetry breaking is kept.
Replicas under 
$\sum^{\prime}$ or $Tr^{\prime}$ are distinct. The $Q^{aa}Q^{bb}$--coupling 
is renormalization group generated as in the metallic quantum spin 
glass, its effects will be discussed in a subsequent section. The
simultaneous
appearance of critical diagonal and off--diagonal field components is
very
unusual for classical thermal spin glass transitions and was so far
known only to occur in special
limits \cite{Moore}.

The tricritical behavior can be extracted from 
a quadratic approximation of the self - consistency equations. Since replica
symmetry breaking is only generated by the $y_4$ - term in the free energy
in the context of a quadratic approximation it is sufficient to consider the
plateau height $q(1) = q_{EA} = q + $ subleading corrections.

\begin{eqnarray}
0&=&g r_g - r_T \delta T^2+6 \delta T \delta\tilde{q}- 
\frac{3}{4}\delta\tilde{q}^2+ 3 q^2
\label{nineteen}\\
0&=&6 q (\delta\tilde{q}-\delta T-q)\label{twenty}
\end{eqnarray}

From (\ref{nineteen}) we get for q=0

\begin{eqnarray}
\delta\tilde{q}_{dis}&=&4(\delta T \pm|\delta T|\sqrt{1+\frac{r_g g}
{12 \delta T^2}-\frac{r_T}{12}})\nonumber\\
&=& 4(\delta T \pm |\delta T| W)
\label{twentyone}
\end{eqnarray}
Only the smaller solution (eg that with the - sign) corresponds to a minimum
of the free energy;
on the  line $\delta \mu = - 3(1/\zeta - \mu_{c3}) \delta T= - 0.1354 \delta 
T$  which is tangent to the critical curves $\mu_{c1}(T)$ and $\mu_{c2}(T)$ 
g vanishes and  $\delta \tilde{q}_{dis}$ is proportional to $\delta T$. The 
statement holds true for a region close to the tangent where $g \sim \delta T
^2$ as well, there 
the radicand is positive for $g<-\frac{12-r_T}{r_g} \delta T^2$.
This result reproduces an expansion of the exact relation $\mu_{c2}(T)$ around
$T=\frac{1}{3}$.

 However, if g is of order $\delta T$ or larger the solution becomes to
leading order $\delta \tilde{q} = \sqrt{\frac{r_g g}{12}}$ and thus displays 
a nonanalytical dependence on temperature and / or chemical potential. This 
type of crossover can also be seen in the free energy

\begin{eqnarray}
f_{dis}&=&\frac{3}{2} |\delta T|^3[4 (sgn(\delta T) - W)(\frac{r_g g}{\delta T^2}
 -r_T)
\label{twentytwo}\\
 & &-48 sgn(\delta T)(sgn(\delta T)-W)^2-16 (sgn(\delta T)-W)^3]\nonumber\\
 &=&|\delta T|^{2-\alpha} {\cal G}(\frac{g}{\delta T^2})\nonumber
\end{eqnarray}

The scaling form allows for the identification of the specific heat exponent
$\alpha =-1$ and the crossover exponent $\phi = 2$. The crossover function
${\cal G}(x)$ is regular for small x and  has the asymptotic form
\begin{equation}
{\cal G}(x) {\approx  \atop x\to\infty} 
x^{\frac{2-\alpha}
{\phi}} {\cal G}_{\infty}+({\rm regular\hspace{.1cm}corrections}). 
\end{equation}
In the tricritical region the
leading singularity in the free energy is given by

\begin{eqnarray}
f_{TCP, sing}&=& \frac{4}{\sqrt{3}}(r_g g)^{\frac{3}{2}}
\label{twentythree}
\end{eqnarray}

For $\delta\mu=0$ we have $g\sim \delta T$ and can read off the tricritical
specific heat exponent $\alpha_3=\frac{1}{2}$ from above.

The next step is the search for ordered solutions of the selfconsistency 
equations. Using $q=\delta\tilde{q}-\delta T$ one solves readily for

\begin{eqnarray}
\delta\tilde{q}_{ordered}=\pm\frac{2}{3}|\delta T|\sqrt{-\frac{r_g g}
{\delta T^2}+r_T-3}
\label{twentyfour}
\end{eqnarray}
The solution exists for $g<-\frac{\delta T^2}{r_g}(r_T-3)$ (see curve d in figure
2). The corresponding replica overlap q is positive for
$q=|\delta T|(-sgn(\delta T)\pm\frac{2}{3}\sqrt{-\frac{r_g g}{\delta T^2}+r_T-3})
>0$. For positive $\delta T$ the "+"- sign needs to be chosen, the condition
$q>0$ is then equivalent to $g<-(\frac{21}{4}-r_T)\frac{\delta T^2}{r_g}$, which
reproduces an expansion of $\mu_{c1}(T)$ (see curve a in figure 2).
Considering now the case $\delta T<0$, we get the condition
$\pm \frac{2}{3}\sqrt{-\frac{r_g g}{\delta T^2}+r_T-3}>-1$.
The "+"  - solution is always o.k., the "-" - solution only in the region 
with $-(\frac{21}{4}-r_T)\frac{\delta T^2}{r_g}<g<-\frac{\delta T^2}{r_g}
(r_T-3)$ (in the region between curve a and d).
The saddle point energy for the physically meaningful "+" - solution is

\begin{eqnarray}
f_{ordered}= - \frac{2}{3}(-\frac{r_g g}{\delta T^2}-\frac{r_T-3}{r_g})
\label{twentyfive}
\end{eqnarray}
The energy difference $f_{ordered}-f_{dis}$ becomes negative for 
$g<-7.9488 \delta T^2$ (thermodynamic transition, shown in curve c in 
figure 2).
In the tricritical region the solution of the selfconsistency equations is

\begin{eqnarray} 
q=\delta\tilde{q}=\frac{2}{3}\sqrt{-r_g g}
\label{twentysix}
\end{eqnarray}

which yields the tricritical order parameter exponent $\beta_3=\frac{1}{2}$
and suggests that $q$ and $\delta \tilde{q}$ act as order parameters 
simultaneously. From the fluctuation Lagrangian one reads off mass squared
proportional to $\delta T^2$ and hence $\gamma_3 = \beta_3 = \alpha_3 = 
\frac{1}{2}$.\\


\subsection{Replica Symmetry Breaking and tricritical Almeida Thouless - 
Line}

In contrast to crystal--field split spin glasses
\cite{MS} the quartic coefficient $y_4$ of our free
energy,
Eq.(\ref{seventeen}), is nonzero and one obtains the Parisi solution 

\begin{eqnarray}
q(x)=\left\{\begin{array}{l}\frac{9}{2y_4}x\quad {\rm for}\quad 0\leq 
x\leq x_1\\ 
q(1)\quad{\rm for}\quad x_1\leq x\leq 1\end{array}\right. 
\end{eqnarray}

The plateau height is found to satisfy 
\begin{equation}
q(1)=\delta\tilde{q}+O(\delta\tilde{q}^2). 
\end{equation}
Consequently, plateau and breakpoint scale like $\sqrt{|\tau|}+O(\tau)$ 
at the $TCP$, while linear $\tau$-dependence is reserved 
to $T_c>T_{c3}$. Adapting the notation of 
\cite{FiSo} we express our result for the irreversible response 
$q(1)-\int^{1}_{0}q(x)\sim |\tau|^{\beta_{\Delta}}$ in terms of the 
exponent $\beta_{\Delta 3}=1$ for $T\rightarrow T_{c3}$ and 
$\beta_{\Delta}=2$ for $T\rightarrow T_c>T_{c3}$.
For the Almeida--Thouless line at tricriticality we find
\begin{equation}
\frac{H^2}{J^2}=\frac{80}{81}(\frac{2}{3}(1-\frac{\mu_{c3}}{J}
tanh(\frac{3\mu_{c3}}{J})))^{3/2}\tau_{AT}^{3/2}+O(\tau_{AT}^2)
\end{equation}
with $\tau_{{\small AT}}\equiv \frac{T_{c3}-T_{AT}(H)}{T_{c3}}$. Hence we 
obtain the critical exponent $\theta_{3}=\frac{4}{3}$ near $T_{c3}$, while 
$\theta=\frac{2}{3}$ for all $T_c>T_{c3}$. 
These values do not 
satisfy the scaling relation $\theta_{3}=
\frac{2}{\beta_{\Delta_3}}$ with 
$\beta_{\Delta_3}=1+(\gamma_3-\alpha_3)/2$. Along the lines described in
\cite{FiSo}, this problem of mean--field exponents will be 
resolved 
below by the renormalization group analysis of the coupling $y_4$ of the 
{\it finite--range and finite--dimensional $ISG_f$}.

\subsection{Stability of the solution in the tricritical region}

A first test for stability of the saddle point solutions is looking at 
the (matrix) of second derivatives of the free energy with respect to q 
and $\delta \tilde{q}$.
\begin{eqnarray}
\frac{\partial^2 f}{\partial \tilde{q}^2}&=&9\delta T - 
\frac{9}{4} \delta \tilde{q} + O(q^2, \delta\tilde{q}^2)
\label{twentyseven}\\
\frac{\partial^2 f}{\partial q^2}&=& 9(-\delta T+ \delta\tilde{q} - 2 q)
\label{twentyeight}
\end{eqnarray}
In the tricritical regime $\delta\tilde{q}$ and $q$ are leading compared
to the temperature deviation. The stability condition $\frac{\partial^2 f}
{\partial q^2} < 0$ is satisfied both in the ordered and in the disordered
regime, whereas $\frac{\partial^2 f}{\partial \tilde{q}^2} $ is 
positive in the paramagnetic phase and negative (e.g. unstable) in the ordered
phase.
One may wonder whether the coupling between diagonal and offdiagonal 
fluctuations cures this
problem or whether there is a new type of instability in addition to the 
well known AT - instability.
. The $\delta Q^{aa}$ - fluctuations 
are vectors in replica space and play
the role of the $\delta m^a$ -fluctuations in the notation of Almeida
and Thouless, the matrix 
entries A, B, C and D
of AT are replaced by (to order $\sqrt{g}$)
\begin{eqnarray}
X&=&\frac{1}{2} \beta^2[1-\frac{1}{2} \beta^2(<(\sigma^a)^4>-
<(\sigma^a)^2>^2)]\nonumber\\
& =& - \frac{27}{4} \delta\tilde{q}
\label{twentynine}\\
Y&=&-\beta^4(<(\sigma^a)^2(\sigma^b)^2>-<(\sigma^a)^2><(\sigma^b)^2>)
\nonumber\\
&=&0
\nonumber\\
V&=&-\beta^4(<(\sigma^a)^3 \sigma^b>-<(\sigma^a)^2><\sigma^a \sigma^b>)
\nonumber\\
&=&-54 q\nonumber\\
W&=&-\beta^4(<(\sigma^a)^2\sigma^b\sigma^c>-<(\sigma^a)^2><\sigma^b>
<\sigma^c>)\nonumber\\
&=&0\nonumber
\end{eqnarray}

P, Q and R control as usually offdiagonal fluctuations and are given by

\begin{eqnarray}
P&=&\beta^2[1-\beta^2(<(\sigma^a)^2(\sigma^b)^2>-<\sigma^a\sigma^b>^2)]
\label{thirty}\\
&=&-54\delta\tilde{q}\nonumber\\
Q&=&-\beta^4(<(\sigma^a)^2\sigma^b\sigma^c>-<\sigma^a\sigma^b><\sigma^a
\sigma^c>)\nonumber\\
&=&-27 q\nonumber\\
R&=&-\beta^4(<\sigma^a\sigma^b\sigma^c\sigma^d>-<\sigma^a\sigma^b>
<\sigma^c\sigma^d>=0\nonumber
\end{eqnarray}

In the limes $n\to 1$ AT obtained five  eigenvalues, one of them
($ P-2Q+R $) is not related to diagonal fluctuations, the other four 
merge in the replica limit to

\begin{eqnarray}
\lambda_{\pm}&=&\frac{1}{2}(X-Y+P-4 Q+3 R\nonumber\\
& &\pm \sqrt{(X-Y-P+4 Q-3 R)^2-8 (V-W)^2}\nonumber\\
&=&\frac{1}{2}(X+P-4 Q\pm\sqrt{(X-P+4 Q)^2-8 V^2}\nonumber\\
&=& \frac{27 q}{8}(7\pm i \sqrt{2^9-3^4})
\label{thirtyone}
\end{eqnarray}

 The real part of these eigenvalues is positive and guarantees stability,
 the logarithm of the partition function stays real in spite of the 
 imaginary part of the eigenvalues, it contains a factor

\begin{eqnarray}
\frac{1}{\sqrt{(\lambda_+\lambda_-) (\lambda_+\lambda_-)^(n-1)}}
\label{thirtytwo}
\end{eqnarray}

which is real in the replica limit.\\
\subsection{Tricritical behavior in the Spin 1 Ising model in transverse 
fields}
The quantum phase transitions at $T=0$ of metallic spin glasses and of the 
transverse field Ising spin $\frac{1}{2}$ glass were recently described in 
a unifying quantum field theoretical way. The essential difference between 
the two being the marginal relevance of the quantum dynamical interaction 
of transverse field models in contrast to their (dangerous) irrelevance
in the metallic case. Similar properties of the phase diagrams of spin 1 
and of the $ISG_f$ suggest a comparison between quantum extensions of 
these two models. The Spin 1 Ising spin glass in a transverse field will
see its thermal tricritical point descend towards zero temperature as the 
transverse field approaches a characteristic value. Again, as in the 
metallic
case, we approximate this value by a Q--static approximation, improving 
this result finally by a generalized Miller--Huse method \cite{MilHu}.\\

\subsection{Renormalization Group Analysis}
\subsubsection{Tricritical Ising Spin Glass}
We performed a 1--loop RG calculation for tricritical fluctuations.
At each RG step the mass of charge fluctuations $\delta Q^{aa}$ was 
shifted away.
Introducing the anomalous dimensions $\tilde{\eta}$ and $\eta$ for diagonal
and offdiagonal fluctuations one finds at one loop level the following
RG relations ($\epsilon = 8 - d$)
\begin{eqnarray}
\frac{dr}{dl}&=&(\frac{d}{2}-11\kappa^2_1+16\kappa_1\kappa_2+6\kappa^2_2)r-
\kappa^2_2\\
\frac{du}{dl}&=&2(1-\kappa_1^2)u-4\kappa_1^2+4\kappa_1\kappa_2,
\nonumber\\
\frac{d\kappa_1}{dl}&=&\frac{\epsilon}{2}\kappa_1+9\kappa_1^3\,
\frac{d\kappa_2}{dl}=(\frac{\epsilon}{2}+6\kappa^2_2-
\kappa_1^2+16\kappa_1\kappa_2)\kappa_2\nonumber\\
\frac{d\kappa_3}{dl}&=&(\frac{\epsilon}{2}+9\kappa^2_2)\kappa_3,\nonumber
\end{eqnarray}
Above $d=8$ the RG flows towards the Gaussian fixed point with
mean field exponents, for $d<8$ there is no perturbatively 
accessible fixed point for real $\kappa's$.
However, a preliminary analysis of the resulting strong coupling problem 
shows that
there exists a solution with positive $\tilde{\eta}$ in contrast to the 
negative anomalous dimension typical of cubic field theories with imaginary
coupling.
The RG for the $DIC$ $y_4$ showed that its long--distance behavior 
is dominated by a $\kappa^4$--contribution (like in \cite{FiSo} but) for 
$d_c^{(u)}=8<d<10$.
This leads to the modified MF exponent 
$\theta_3=8/(d-4)$, which satisfies the scaling relation 
$\theta_3=2/\beta_{\Delta_3}$ in $d_{c3}^{(u)}=8$ 
and reduces to the MF--result 
in $10$ dimensions.
The dimensional shift by $2$ in comparison with \cite{FiSo} is due to 
coupling $t$. 


\subsubsection{Metallic Quantum Spin Glass}

In the theory of quantum phase transitions time--dependent fluctuations
are treated on an equal footing with spatial fluctuations 
\cite{SRO}. While
for the Lagrangian (\ref{seventeen}) it was sufficient to consider only 
the $\omega = 0$ - component of the Q - fields, in the quantum case
all low energy fields must be kept as they are coupled via the quantum
mechanical interaction $u\int d\tau (Q^{aa})^2$. However, the value 
$z = 4 $ of the dynamical
critical exponent renders the u  - coupling dangerously irrelevant
and allows for a perturbative mapping of the critical theory on 
a classical problem, the Pseudo - Yang - Lee edge singularity. 
This field theory has only one cubic coupling which corresponds to
$\kappa_1$ in eq(\ref{seventeen}). The comparison of the metallic 
quantum case with the thermal tricritical theory allows one to understand
the nature of the strong coupling RG fixed point: whereas the spin 
fluctuations in the thermal second order regime are governed by a 
perturbatively accessible fixed point, the TCP and the quantum case
are characterized by the combination of charge and spin fluctuations 
and a corresponding strong 
coupling problem.


\section{The paramagnetic phase: Multiple  low temperature
  saddle--point solutions}
Complications due to variation of the chemical potential can also be seen
outside the spin glass phase of the fermionic Ising spin glass.
We have discussed (see also Fig.(\ref{muc2})) the unusual critical line 
$\mu_{c2}(T)$, which crosses the spin glass regime and, passing 
through the thermal tricritical point, stretches into 
the regime up to arbitrarily large $\mu$. It was seen that in the regime of
large particle pressure 
the spin density becomes small and spin glass order is apparently absent. 
Despite the presence of the $\mu_{c2}(T)$-- or, if one wish, 
$T_{c2}(\mu)$--line, the absence  of magnetic phase transitions caused
by the totally frustrated magnetic interaction is predicted by the
replica
symmetric ordered solution eq.(\ref{fifteen}). 
The multitude of paramagnetic solutions below the $T_{c2}(\mu)$--line
and a possible metastability in finite range models is discussed in
this section.
We take advantage of the fact that replica permutation symmetry will
play no role. We consider however possible that spontaneous vector
replica symmetry breaking \cite{dotsmez} will be involved.
In the paramagnetic regime one finds three solutions for $\tilde{q}$, and 
hence for the susceptibility $\chi=\beta\tilde{q}$ too, at temperatures 
below $T_{c2}(\mu)$. This temperature is given by the inversion of
Eq.(\ref{eleven}) and is also shown in Fig.(5).\\
At low temperatures these solutions are well approximated by
\begin{eqnarray}
\tilde{q}\hspace{.1cm}|_{_{_{T<<T_{c2}}}}=\left\{ \begin{array}{l}1\\
2\mu/\beta+O(\beta^{-2}ln\beta)\\
2 e^{-\beta\mu}\end{array}\right.
\end{eqnarray}
These saddle--point solutions for $\tilde{q}$ and the 
corresponding ones for the linear susceptibility are displayed for 
a few characteristic values of the chemical potential $\mu$, starting from
the tricritical value $\mu_{c3}=0.96125 J$. The magnetically saturated 
solution ($\tilde{q}=1$) which would imply a Curie susceptibility $\chi\sim 
\beta$ is unstable with respect to spin glass order for all
temperature. 
The high temperature solution, 
which decays exponentially at low temperatures and leads to
a vanishing susceptibility in the zero temperature limit is the
physical
paramagnetic solution in the whole temperature range. It is not clear
however
whether a replica symmetry broken stable solution with a possibly
lower
free energy exists due to the presence of exceptionally strong bonds.
The linearly decaying solution is a maximum of the free energy at all 
temperatures.
For large values of the chemical potential the physical solution 
$\tilde{q}=2 e^{-\beta\mu}$ has a much larger free energy than the
Curie type solution $\tilde{q}=1$. The two free energy minima are
separated
only by a very small barrier, hence we expect fluctuations to play a
major role in finite range models. The instability of the lower
minimum
with respect to spin glass order makes it
conceivable that due to Griffiths effects a replica symmetry broken
ordered solution
persists in the region of very low T for all values of the chemical
potential.
\begin{figure}
\epsfig{file=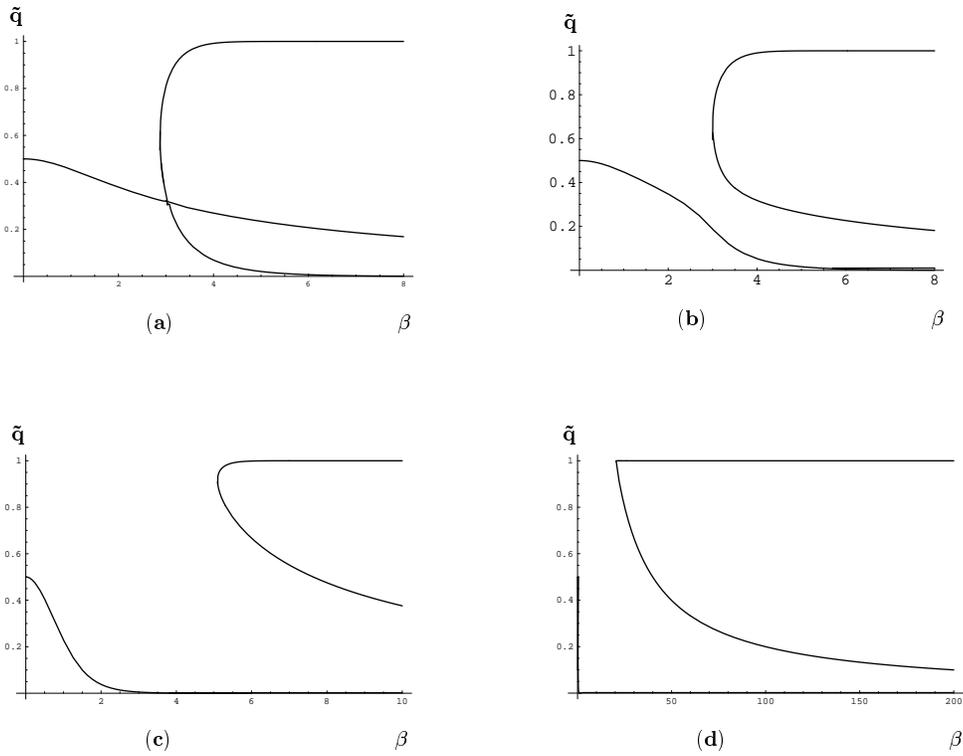,width=10cm,angle=270}
\caption{Spin autocorrelation function shown for various chemical
potentials $\mu=.96125 J (a), J (b), 2 J (c),$ and $10 J (d)$ with three 
low temperature solutions of the paramagnetic saddle-point equation: 
the lowest one (indistinguishable over a wide range from the $\beta$--axis 
on the given scale) is the only stable minimum of the free energy.} 
\end{figure} 
\begin{figure}
\epsfig{file=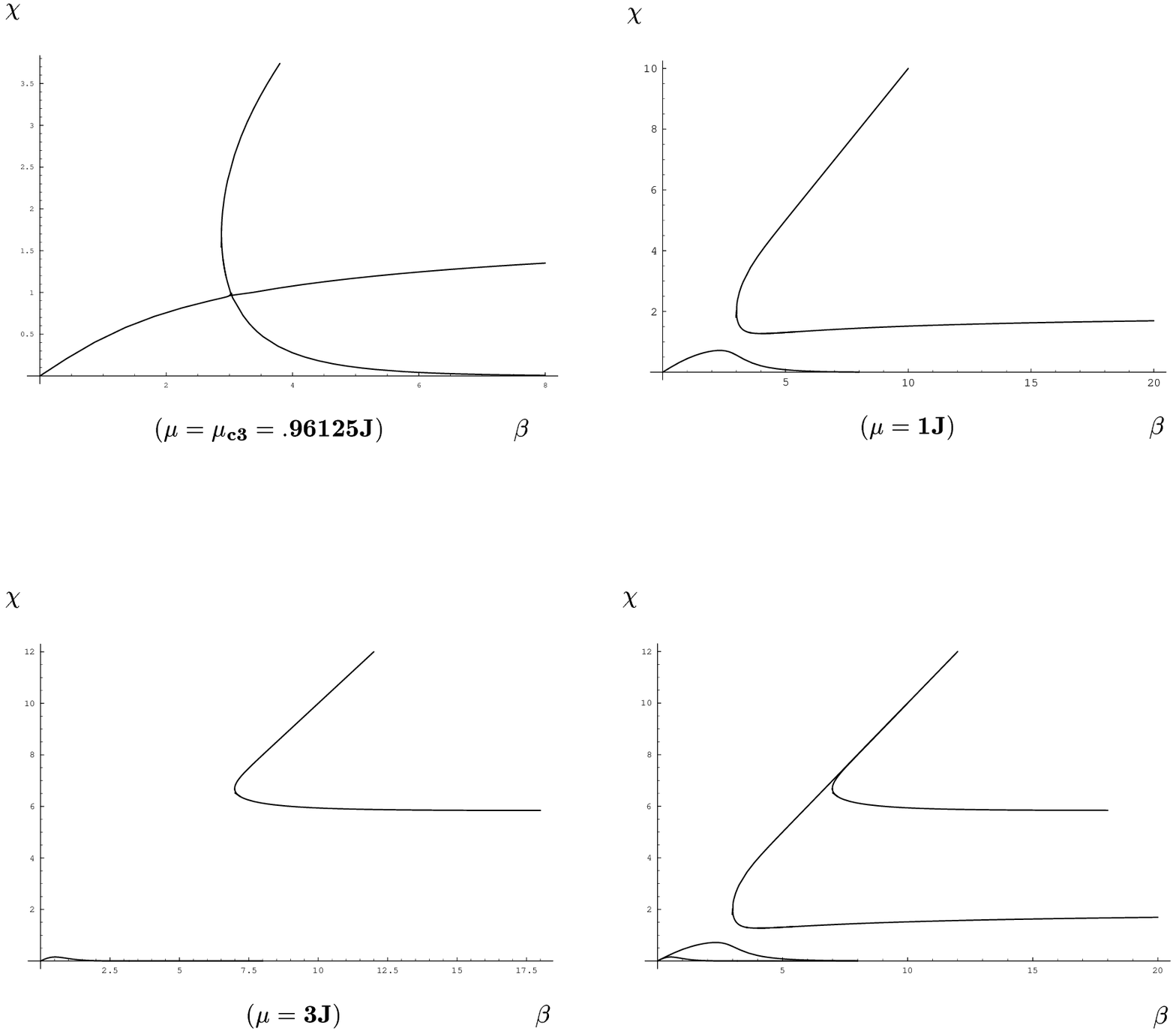,width=15cm,angle=270}
\caption{Linear susceptibility corresponding to the three solutions
of the paramagnetic saddle--point equation, 
given for $\mu=1$, $\mu=3$, which are joined in the fourth figure (bottom
right); the first one (top left) shows the
solution near the tricritical chemical potential for comparison.
The (sole) high temperature solution remains realized at all
temperatures.}
\end{figure}
\begin{figure}
\epsfig{file=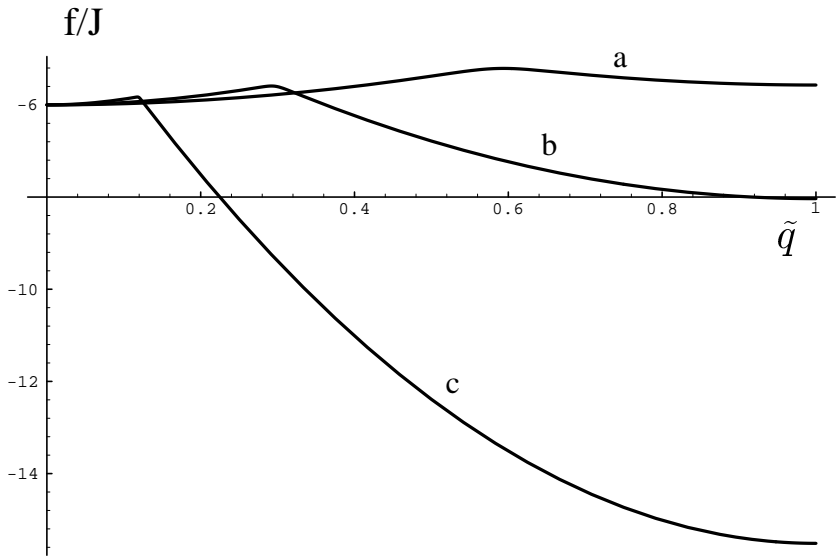,width=8cm,angle=270}
\caption{Paramagnetic free energy as a function of $\tilde{q}$ for
  $\mu =3 J$ and $\beta =10/J$ (curve a),$\beta=20/J$ (b), and $\beta=50/J$
  (c).}
\label{frei}
\end{figure}


\section{ Relations with the Blume-Emery-Griffiths model for
$He^3-He^4$ mixtures; universal mean field tricritical
point}
For classical spin systems there exists a variety of
different ways to describe diluted systems. Spin 1
models, where $S_i=0$ may be considered to describe an
empty site i, provide one example. A beautiful case was
formulated by Syozi, who extended Onsager's solution
to the twodimensional diluted Ising ferromagnet.
Another example is given by the Blume-Emery-Griffith
model intended to describe $He^3-He^4$ mixtures, where
the $S_i=0$ state corresponds to a $He^3$-atom
on site i. Disorder effects added to the original
BEG model have raised considerable interest recently
\cite{FalBer}. But it is in fact the clean
BEG model that shows the perhaps accidental but in any
case most eyecatching resemblance with part of the
fermionic Ising spin glass tricritical behavior. There
is a surprisingly identical tricritical temperature
$T_{c3}=\frac{1}{3}J$ of the clean BEG without
$S^2S^2$-interaction (ie K=0) on one hand and of the
$ISG_f$ on the other (the meaning of J being n.n.
$J_{ij}$ for the BEG and standard deviation of $J_{ij}$ in case
of the spin glass). Another surprising fact is however
that it is the $K\neq 0$ BEG-equation for $<S^2>$ ($=1-x$
in \cite{BEG}) which can be mapped onto the
$ISG_f$-equation for $<(S^a)^2>\sim \tilde{q}$ in the
disordered phase. Reconsidering the $\tilde{q}$-equation
for $H=m=q=0$ from the selfconsistent $ISG_f$-equations
given above
\begin{equation}
\tilde{q}=1/[1+cosh(\beta\mu)exp(-\frac{1}{2}\beta^2
J^2\tilde{q})]
\end{equation}
and comparing it with the BEG-equation for $m=0$ which
reads\\
\begin{equation}
1-x=1/[1+\frac{1}{2}exp(\beta\Delta)exp(-\beta K(1-x))]
\end{equation}
one finds an identical structure in the variables $\tilde{q}$ and $1-x$.

\section{Phase Separation}

Up to now the tricritical phenomenon and the first order transition from
paramagnet to spin glass have mainly been discussed for given chemical 
potential. In this picture a thermodynamical first order transition is
observed for $\delta \mu = - 0.1355 \delta T - 7.9488 \delta T^2$. On
this line the free energies of paramagnetic and ordered solution are the same
but the two solutions correspond to different fermion fillings of the system.
This can be seen from the following consideration: On the first order line
we have $\delta \tilde{q}_{dis} = - 6.2649 |\delta T|$ and 
$\delta \tilde{q}_{ordered} = 1.5133 |\delta T|$. Via the relation

\begin{eqnarray}
\delta \nu = - \delta \tilde{q}  \tanh(\beta_{c3} \mu_{c3}) - \delta T
\frac{1}{2 e^3}
\end{eqnarray}

the phase diagram in $(\nu , T)$ space in the neighborhood of the TCP is 
defined. $\delta \nu_1 = 6.2522 |\delta T|$ limits the ordered region,
$ \delta \nu_2 = - 1.4779 |\delta T|$ the disordered region. For 
intermediate
fillings $\delta \nu_1 < \delta \nu < \delta \nu_2$ both phases coexist. 

\section{ The Fermionic Ising Chain}

The remarkable similarities between the phase diagrams of the BEG--model 
without disorder and the one of the fermionic Ising spin glass can be 
taken as indicative for the fact that rather dilution than disorder is the 
source of the tricritical crossover from continuous to discontinuous phase 
transitions. Moreover the RG theory showed that the tricritical structure 
is also not very sensitive to spatial dimension. These impressions and the 
fact that mapping of fermionic spin glass transitions to Pseudo Yang Lee 
edge singularities have been proved may be enough motivation for an 
analysis of onedimensional cases. \\
While it is complicated to solve 1D 
fermionic Ising spin glasses ( no matter whether $d=1$ is below the lower 
critical dimension or above due to long range interaction)
exactly, the clean fermionic Ising chain offers simple exact solutions. 
These solutions discussed here will provide some insight into the role of 
the chemical potential and moreover generalize known results into the 
complex $\mu$--plane.\\
Yang and Lee \cite{yanglee} derived the distribution of zeroes of the 
partition function of finite and infinite Ising chains within the complex 
magnetic field plane. In fermionic Ising systems the chemical potential 
can be seen as complementary to the magnetic field. Stimulated by the 
representation of conventional Ising chains by fermionic ones with special 
imaginary chemical potential, one may wish to extend the Yang Lee 
analysis to a fourdimensional space $((Re\mu),Im(\mu),Re(h),Im(h))$.\\
The transfer matrix ${\bf T_f}$ of the fermionic Ising chain in a 
(complex) field $h$ and with (complex) chemical potential $\mu$ reads
\begin{eqnarray} T_f& =& e^{\beta\mu}\left( 
\begin{array}{*{4}{c}} 
e^{\beta\mu} & 1 & 
e^{\frac{1}{2}\beta(\mu +h)} & e^{\frac{1}{2}\beta(\mu-h)} \\
1 & e^{-\beta\mu} & e^{\frac{1}{2}\beta(h-\mu)} & 
e^{-\frac{1}{2}\beta(\mu+h)}\\
e^{\frac{1}{2}\beta(\mu+h)} & e^{\frac{1}{2}\beta(h-\mu)} & 
e^{\beta(J+h)} & e^{-\beta J}\\
e^{\frac{1}{2}\beta(\mu-h)} & e^{-\frac{1}{2}\beta(\mu+h)} & e^{-\beta 
J} & e^{\beta(J-h)}
\end{array} \right)\nonumber\\
&& 
\end{eqnarray}
while the corresponding one of the $S=\pm1$--chain is given 
by
\begin{eqnarray} T_s =  \left( \begin{array}{cc} e^{\beta(J+h)} & 
e^{-\beta J} \\ e^{-\beta J} & e^{\beta(J-h)} \end{array} \right). 
\end{eqnarray}
The transfer matrices and their eigenvalues don't map onto each other at
$\mu=i\frac{\pi}{2}T$, but the partition function does according to\\
\begin{equation}
Z_f^{(N)}=Tr T_f^N =(2i)^NZ_s^{(s)}
\end{equation}
for any number N of sites.\\
The largest eigenvalue determines the free energy of the infinite chain,
while the second largest is required in addition to determine the 
correlation length. The largest eigenvalue results in general from a 
cubic equation resulting in a lengthy result for the free energy.
For several purposes it is sufficient to know the eigenvalues for
$(h=0,\mu)$, $\mu$ arbitrary complex. The half--filled case 
$(\mu=0, h)$, h arbitrary complex, may also be considered separately.\\
The eigenvalues for $h=0$ are found as\\
\begin{eqnarray}
\nonumber
\lambda_{\pm}&=&e^{\beta\mu}[ch(\beta\mu)+ch(\beta
J)\\
& &\pm\sqrt{(ch(\beta\mu)+ch(\beta J))^2+4 ch(\beta\mu)(1-ch(\beta
J))}]\nonumber\\
\lambda_0&=&0\quad,\quad  \lambda_1=2e^{\beta\mu}sh(\beta J).
\end{eqnarray}
The correlation length is given by\\
\begin{equation}
\xi = 1/ln(\frac{\lambda_+}{\lambda_1}).
\end{equation}
In the $T\rightarrow 0$--limit a transition (formally) arises at $\mu=J$ 
and due to the properties
\begin{eqnarray}
\lambda_1&\sim& exp(\beta(\mu +J))\\
\lambda_+&\sim& 
\left\{ \begin{array}{cc} exp(\beta(\mu+J)),\quad \mu < J\\ 
exp(2\beta\mu), \quad \mu > J \end{array} 
\right.\nonumber
\end{eqnarray}
Thus the correlation length diverges only for $\mu<J$, which is 
comprehensible since the energy required for adding a fermion is larger 
than the gain from a magnetic bond if $\mu>J$. One obtains 
\begin{eqnarray}
\xi\sim \left\{ \begin{array}{c} exp(\beta(2 J -\mu)) \quad ,0 < \mu < J \\ 
\exp(\beta J/2 \quad , 
\mu = J.
\end{array} \right.
\end{eqnarray}
while  $\xi\sim T/(\mu - J)$ for $\mu > J$.
Taking the derivative of the free energy $f=-T ln \lambda_+$ w.r.t. the 
chemical potential yields the filling factor
showing that
in the zero temperature limit the system is completely 
filled provided $\mu$ is positive (empty for negative $\mu$). Thus there
is no physical $T=0$--transition of this simple system. The correlation 
length diverges in the $T\rightarrow 0$ limit for all fillings $\nu$.\\
The zero--field partition function shows that Yang--Lee zeroes approach
$|\mu|=\pm J$ for $T\rightarrow 0$. This means that for 
$(h=0,T=0)$ nonanalytic behavior (as a function of the real chemical 
potential) can only occur at the values $\mu =\pm J$. \\
It is instructive to consider $N=2$ explicitly, which yields
\begin{eqnarray}
Z^{N=2}_f&=&4e^{2\beta\mu}[(ch(\beta\mu)+ch(\beta J))^2\\
&& + ch^2(\beta h)(e^{2\beta J}-1) - sh(2\beta J)]\nonumber
\end{eqnarray}
This almost trivial case already shows zeroes at
\begin{equation}
\mu_0=\pm(J+(\frac{1}{2}ln2\pm i\frac{\pi}{2})T)
\end{equation}
while allowing for finite complex magnetic field the first zero different
from $\pm J$ in the $T\rightarrow 0$--limit becomes possible with
\begin{equation}
\mu =\pm(i\frac{\pi}{4}+ 2 im\pi)T = \mu + i m\pi T .
\end{equation}
More zeroes occur on the $T=0$--axis as N is increased. In the 
$N\rightarrow 0$--limit a density function is expected in accordance with
the divergent behavior of the correlation length for any $\mu$. 
Its evaluation is beyond the scope of this paper.\\

\section{Mapping the twodimensional Ising model with imaginary 
magnetic field $h=\frac{i\pi}{2}T$ into fermionic space}
The complementary role of complex magnetic field and complex chemical potential
can nicely be seen by recalling that the 2d Ising model with nonreal magnetic 
field had only been solved exactly at the special value $H=i\frac{1}{2}\pi T$.
This value corresponds to $\mu=i\frac{1}{2}\pi T$, which maps the fermionic
Ising model onto the one above. Thus the exact solution of the 2d fermionic 
Ising model with $\mu=H=i\frac{1}{2}\pi T$ is known. Moreover this special
model maps onto an interacting model which describes the interaction of spinless 
fermions with a special species obeying bare Bose statistics but carrying
along the minus signs of fermion interactions. 
The Hamiltonian of this model can be written
\begin{equation}
H=-\sum_{ij}J_{ij}\sigma_i^z\sigma_j^z-\mu\sum_i 
(n_{i\uparrow}+n_{i\downarrow})-h\sum_i \sigma_i^z
\end{equation}
with $\mu=h=i\frac{\pi}{2}T$. While this value of $\mu$ provides the desired
map between 2--state-- and 4--state--per--site--space for any magnetic field, 
the special value considered here reduces to zero the imaginary field of one 
fermionic species, while the other acquires a shift equal to the distance 
between Bose-- and Fermi--Matsubara energies. \\
Setting $c=a_{\uparrow}$ and 
$d=a_{\downarrow}$ the Hamiltonian reads
\begin{eqnarray}
H&=&-\sum_{ij}J_{ij}[c^{\dagger}_i c_i c^{\dagger}_j c_j
- c^{\dagger}_i c_i d^{\dagger}_j d_j - d_i^{\dagger}d_i c_j^{\dagger}c_j 
+
d^{\dagger}_i d_i d^{\dagger}_j d_j] -\nonumber\\
& & i\pi T\sum_i d^{\dagger}_i d_i.
\end{eqnarray}
The effect of the new imaginary chemical potential $i\pi T$ acting only on
the d--fermions can both be imagined in standard many body diagram 
theory and within the fermionic path integral representation. In both 
formalisms the fermionic Matsubara energies of the d--fermions become 
bosonic ones. In other words d--propagators become bosonic while 
c--propagators remain (unchanged) fermionic. All interactions (vertices)
between c--fermions and d--particles remain fermionic in character, ie 
they carry
the minus--signs due to the anticommutation rules of the operators c and 
d. In the fermionic path integral representation Grassmann--fields are 
indispensable for both c-- and d--operators. Denoting them by 
$\psi_c(\tau)$ and by $\psi_d(\tau)$ respectively, the important change 
due to the $i\pi T \hat{n}_d$--term, which we absorb by the phase 
transformation $exp(i\pi \tau)\psi(\tau)=\tilde{\psi}(\tau)$ in the 
new anticommuting fields $\tilde{\psi}$, occurs in the 
unusual periodicity of $\psi_d(\tau)$ on the imaginary time axis:
\begin{equation}
\tilde{\psi}_d(\tau) = \tilde{\psi}_d(\tau +\beta)
\end{equation}
while $\psi_c(\tau)=-\psi_c(\tau + \beta)$ retain the fermionic 
antiperiodicity. The periodic behavior of the $\psi_d$'s purports of 
course into the bosonic propagator form. \\
So far the mapping has been exact. In a perturbatively exact way one can 
even go beyond the above conclusion by stating that the 2D Ising model
for this special imaginary magnetic field maps onto a coupled Fermi--Bose 
system (c--d) with the speciality that any d--loop contributes  
additional $(-1)$--factors. Despite the speciality of identical coupling
constants $J_{ij}$ between all species this system is physical in the 
sense that there are no imaginary fields, but the mapping of the known
solution of the 2d Ising model discussed above is of order by order 
character.        \\

\section{Acknowledgement}
Research presented in this paper (II) and in the preceding paper (I) was 
supported
by the Deutsche Forschungsgemeinschaft under project Op28/5--1,
by the SFB410 of the DFG, and by 
the Deutsche Studienstiftung.\\
We benefitted from hints by C. De Dominicis, B. Derrida, J. Hertz, D. 
Sherrington, from continued support by H.A. Weidenm\"uller,  and from 
critical comments of our coworkers at W\"urzburg 
H. Feldmann and M. Rehker.


\end{document}